# AN STPA-BASED APPROACH FOR SYSTEMATIC SECURITY ANALYSIS OF IN-VEHICLE DIAGNOSTIC AND SOFTWARE UPDATE SYSTEMS

**Jinghua Yu [1]   Stefan Wagner [2]   Feng Luo [3]**

[1,3] *Tongji University, School of Automotive Studies,
Caoan Highway 4800, 201804 Shanghai, China ([1]yujinghua@outlook.com, [3]luo_feng@tongji.edu.cn)*
[2] *University of Stuttgart, Institute of Software Technology,
Universitaetsstrasse 38, 70569 Stuttgart, Germany (stefan.wagner@iste.uni-stuttgart.de)*

**ABSTRACT**: The in-vehicle diagnostic and software update system, which supports remote diagnostic and Over-The-Air (OTA) software updates, is a critical attack goal in automobiles. Adversaries can inject malicious software into vehicles or steal sensitive information through communication channels. Therefore, security analysis, which identifies potential security issues, needs to be conducted in system design. However, existing security analyses of in-vehicle systems are threat-oriented, which start with threat identification and assess risks by brainstorming. In this paper, a system-oriented approach is proposed on the basis of the System-Theoretic Process Analysis (STPA). The proposed approach extends the original STPA from the perspective of data flows and is applicable for information-flow-based systems. Besides, we propose a general model for in-vehicle diagnostic and software update systems and use it to establish a security analysis guideline. In comparison with threat-oriented approaches, the proposed approach shifts from focusing on threats to system vulnerabilities and seems to be efficient to prevent the system from known or even unknown threats. Furthermore, as an extension of the STPA, which has been proven to be applicable to high level designs, the proposed approach can be well integrated into high-level analyses and perform co-design in different disciplines within a unified STPA framework.

## 1. Introduction

Connected vehicles are a big trend in the automotive industry. The Vehicle-to-everything (V2X) innovations provide better traffic safety and management [1] by increasing interactions between vehicles and the outside world. Yet, this leads to rising security challenges. The in-vehicle diagnostic and software update system, which is an essential part of modern vehicles that supports remote diagnostic and OTA firmware or configuration updates, is a critical attack goal in automobiles. Adversaries can install malicious software into vehicles or steal sensitive information like business secrets or personal data and cause property or life losses.

Therefore, security needs to be considered in the system lifecycle. Security analysis, as a key step in early design, discusses potential security issues and derives related items, like threats, risk assessment and constraints, to guide secure design.

The purpose of this research is to propose a suitable approach for systematic security analysis of in-vehicle network systems and provide the analysis guideline for in-vehicle diagnostic and software update systems. The outcomes of the analysis, containing traceable system vulnerabilities and security constraints, would be the inputs of the next security design phase and helpful to migrate risks of being attacked.

The remaining paper is structured as follows. In section 2, we introduce the state of the art with identified gaps and our contributions. In section 3, a system model is proposed for the target system. Section 4 describes the optimized approach with analysis guideline and an example case. Section 5 discusses the comparison between two sorts of approaches and presents the highlights and outline of this research. Finally, we conclude this paper and propose future work in section 6.

## 2. Related Work and Our Contributions

### 2.1. Security analysis of vehicle diagnostic and software update systems

Since diagnostics and SW update are essential functions in vehicles, research has been conducted to analyze security issues of these systems. L. C. Wei [2] performed analyses of three OTA software update scenarios by using two different approaches, achieved security requirements and concluded the merits and demerits of different approaches. C. Schmittner et al. [3] analyzed an OTA system and identified the threat mode, cause and effect, misuse or attacker cases. However, these analyses focus on issues at a high level. The vehicle, together with other entities like manufacturers and governments, is regarded as a component in such a socio-technical system. Yet, crucial information from the lower levels of in-vehicle parts are not part of the analyses.

As to the in-vehicle aspect, the E-safety Vehicle Intrusion Protected Applications (EVITA) project [4] presented its security analysis of various use cases by using an asset-oriented approach. It started with identifying system assets and threats, evaluated risks by the attack tree and achieved security requirements. K. Shanmugam et al. [5] also analyzed security issues of automobile software update systems by defining the threat model with attack goals and capabilities first. J. Lindberg [6] identified weaknesses of the Diagnostic communication over Internet Protocol (DoIP) protocols by using the method defined in the NIST SP 800-30 standard. R. B. Gujanatti [7] presented an analysis of four DoIP communication scenarios based on two attacker types (active and passive) and general security attributes, like data origin authenticity, integrity and confidentiality. All these analyses started from the threat model with general security requirements.

To sum up, the existing system-oriented analyses all focus on high-level system views in which the vehicle is regarded as a single component, while the analyses for in-vehicle systems are threat-oriented and achieve security requirements largely depending on brainstorming.

### 2.2. Security Analysis Approaches

The analysis approaches can be categorized into two classes. The first one is the component-based class, containing the Security Aware Hazard Analysis and Risk Assessment (SAHARA) and the Failure Mode, Vulnerabilities and Effect Analysis (FMVEA), which extend existing safety analysis techniques into security fields. These approaches start with a system decomposition and focus on component failures. The other class is the system-based one, containing the Combined Harm Assessment of Safety and Security for Information System (CHASSIS) and the STPA-based approaches, which start on a higher system level and put more emphasis on interactions between system components. [8].

STPA is a safety analysis technique based on the System-Theoretic Accident Model and Processes (STAMP) model. Other than component failures, it considers hazards caused by unsafe control actions [9]. STPA-Sec is a security extension of STPA sharing a similar framework but extending it by security considerations [10]. Later, limitations like the lack of types of non-safety related losses and no guidance on identifying causal factors in the security domains were addressed by I. Friedberg et al. [11]. They proposed a new unified approach called STPA-SafeSec. F. Warg [12] also identified two limitations and proposed an optimized extension of STPA-Sec. Furthermore, STPA-Priv was proposed to deal with privacy issues by introducing privacy-related concepts and consequence into the STPA framework [13][14].

Some comparisons between approaches have been made. C. Schmittner [3] presented a case study of safety and security co-analysis for automotive cyber-physical systems by using FMEVA and CHASSIS. CHASSIS is suitable for early engineering stage at a high level and largely relies on expert knowledge, while FMEVA is better for later phases with more system details and highly depends on system models and available information. L. C. Wei [15] compared the cybersecurity analyses of autonomous vehicles by using STPA-Sec and CHASSIS. STPA-Sec has strengths in analysis from socio-technical aspects and more focuses on top-down mechanisms, while CHASSIS is based on information flows and care about possible activities of misusers or attackers.

### 2.3. Gaps and Contributions

Three gaps are identified. Firstly, all security analyses of in-vehicle systems are threat-oriented, which start with asset and threat identification by brainstorming. No system-oriented analysis exists. Secondly, the existing STPA approaches are based on system control structures, which do not fit well to information-flow aspects of systems. Thirdly, no general system models of in-vehicle network systems exist. Most analyses were performed on particular example cases and cannot be general guides.

To fill the gaps, we propose an optimized STPA-based approach to perform a system-oriented security analysis for information-flow-based systems. Besides, we propose a system model of in-vehicle diagnostic and software update systems based on which the security analysis guideline is presented.

## 3. System Model

### 3.1. General Description

The vehicle diagnostic and software update system is a data transmission system based on in-vehicle networks like Controller Area Networks (CAN) and Automotive Ethernet (AE). The system functions can be categorized into two classes – writing and reading. The writing class includes reprogramming or reconfiguring Electronic Control Units (ECU) or triggering pre-defined control routines, and the reading class refers to getting stored information, Diagnostic Trouble Codes (DTC) and working states from vehicles.

To achieve system functions, the in-vehicle system needs to be linked with outside entities, like manufacturer clouds or handheld devices, through various communication channels like wired cable, local wireless channels or the Internet. The overall working environment is shown in Figure 1.

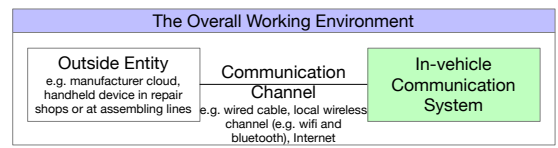

Figure 1 Overall Working Environment

### 3.2. Architecture

The in-vehicle communication system consists of three kinds of components. The vehicle interface is the passageway between the system in vehicles and the outside world. The On-Board Diagnostic (OBD) interfaces and Telematic Control Units (TCU) are common interfaces in modern automobiles. Network components like gateways, cables and switches compose communication channels in vehicles. The end device plays the role as a service provider, which can be an ECU, a sensor or an actuator. The general system architecture is shown in Figure 2.

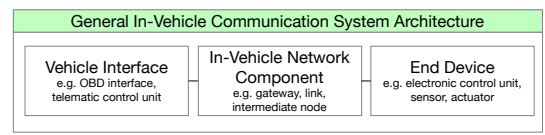

Figure 2 General System Architecture

### 3.3. Working Situation

Who operates the system? When and where is the system used? How does the system work? Various working situations affect the outcomes of security analyses. To achieve unified and accurate descriptions of use cases, working situations are categorized and defined in Table 1.

Table 1 Working Situation Category

| Class | Item |
|---|---|
| Operation Type | Synchronous |
| | Asynchronous |
| Working Location | Manufacturer place |
| | Anywhere else |
| Connection Type | Local |
| | Remote |

'Synchronous' or 'asynchronous' means that data exchange between vehicles and the outside is performed at the same or different time [16]. 'Manufacturer place' refers to assembling lines, repair stations in factories or 4S shops. 'Local' means connections in Local Area Network (LAN) within a limited area (e.g. via wired or wireless channels), while 'remote' represents connections in Wide Area Network (WAN) (e.g. via the Internet).

A combination of functions, architecture and working situations constructs a particular use case, which should be clearly defined before the analysis.

## 4. Methodology

### 4.1. Overview

The proposed analysis steps are listed as follows.

- Step1 – Analyze System-level Concepts: Identify system-level losses and hazards.
- Step 2 – Construct Functional Interaction Structure (FIS): Identify system functions and responsibilities, and then construct the system FIS.
- Step3 – Analyze Insecure Factors: Identify Insecure Function Behaviors (IFB) and Loss Scenarios (LS), and finally derive system security constraints.

The main difference between the proposed approach and the original STPA is to use FIS, instead of the Functional Control Structure (FCS), since it is hard to derive control structures in information-flow-intensive systems. The new FIS is suitable for information systems at both system and component levels and can show physical locations of functions. Besides, new instructors (e.g 'calling behaviors' in Table 6) are proposed for LS identification.

The basic element of the FIS is 'Function' (shown in Figure 3). A function works based on inputs and its algorithm and produces outputs. The function behaviors are affected by the environment. Inputs and outputs, instead of control actions and feedbacks in FCSs, are interaction behaviors in the FIS.

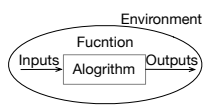

Figure 3 Function - Basic Element in FISs

### 4.2. Analysis Guideline

A guideline for analyzing the in-vehicle communication system is presented in this section. The scope of this analysis only includes information security issues. Therefore, we assume that the system can work properly without intended external disturbances. In other words, system development errors and hardware random failures are out of the scope of this analysis.

#### 4.2.1. Step 1 - Analyze System-level Concept

Firstly, high-level losses are identified in Table 2 according to definitions of safety and security in various industry standards.

Table 2 Identified Losses at High Level

| Index | Loss | Description |
|---|---|---|
| L-1 | Loss of life or cause injury to life | Human and animal life in or outside vehicles |
| L-2 | Loss of physical property | Physical objects belonging to stakeholders (e.g. vehicles, components etc.) |
| L-3 | Loss of non-physical property | Virtual property belonging to stakeholders (e.g. sensitive information, user satisfaction etc.) |
| L-4 | Loss of environment | Natural or artificial world environments |

Then, system-level hazards are identified based on proper security models like the CIA (Confidentiality, Integrity and Availability) Triad. Identified hazards are listed in Table 3 with linked losses.

Table 3 Identified System-level Hazards

| Index | Hazard | Linked Loss |
|---|---|---|
| H-1 | System leaks sensitive information. | L-3 |
| H-2 | System uses intended modified data without detected. | L-1/2/3/4 |
| H-3 | System fails to accomplish missions. | L-3 |
| H-4 | System works with unauthenticated devices. | L-1/2/3/4 |

#### 4.2.2. Step 2 - Construct FIS

In this step, the target system is analyzed and the FIS is created.

Based on the proposed system model, the vehicle interface, as the gate of in-vehicle systems, is responsible for data check, transforming and transmission. The network links include any intermediate entities in networks, whose responsibility is to transmit data. The end device is the communication target in the system, which are responsible for data check, request decapsulation, response encapsulation and providing services. Then, the FIS at the system level is drawn as in Figure 4.

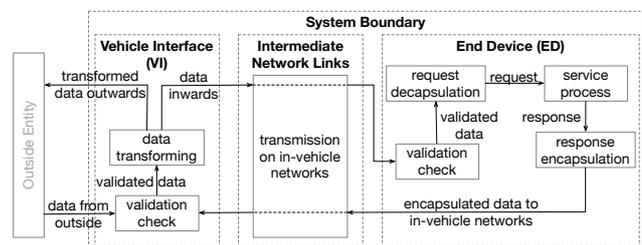

Figure 4 FIS of the Target System

#### 4.2.3. Step 3 - Analyze Insecure Factors

The first sub-step is to identify IFBs, which are behaviors of a function that may cause insecure scenarios and lead to hazards in a particular context like a worst-case environment. Similar to the methods in the STPA approach, three sorts of situations are considered and listed in Table 4 with identification instructors. 17 identified IFBs are then presented in Table 5.

Table 4 Identification Instructors for IFBs

| Not Providing Causes Hazard (NPCH) | Providing Causes Hazard (PCH) | Timing Issue (TI) |
|---|---|---|
| - Not called<br>- Called but not executed successfully | - With incorrect data input<br>- With improper algorithm<br>- With information leakage risk | - Violate time limit |

Table 5 IFBs of the Target System

| Function | NPCH | PCH | TI |
|---|---|---|---|
| Data Check (in VI[1] & ED[2]) | IFB-1 | IFB-2/3 | IFB-4 |
| Data Transforming (in VI), Request Decapsulation & Response Encapsulation (in ED) | / | IFB-5/6 | IFB-7 |
| Data Transmission (on links) | IFB-8 | IFB-9/10/11 | IFB-12 |
| Service Process (in ED) | IFB-13 | IFB-14/15/16 | IFB-17 |
| IFB-n Description with linked hazards ||||

IFB-1: Data check is bypassed and outputs a fake OK. [H-2/4]
IFB-2: Unauthenticated or modified data are not detected.[ H-2/4]
IFB-3: Correct data are input, but outputs a fake NOK. [H-3]
IFB-4: Data check process takes too long and leads to violation of timing. [H-3]
IFB-5: Unauthenticated or modified data are transformed, decapsulated or encapsulated. [H-2/4]
IFB-6: Data are transformed, decapsulated or encapsulated by using malicious algorithms, which may cause insecure behaviors (e.g. modify original information). [H-2/3/4]
IFB-7: Data transformation, decapsulation or encapsulation takes too long and lead to the violation of timing. [H-3]
IFB-8: Data fail to be transmitted to networks. [H-3]
IFB-9: Unauthenticated or modified data are transmitted. [H-2/4]
IFB-10: Data are modified during the transmission. [H-2]
IFB-11: Data are transmitted with information leakage risks. [H-1]
IFB-12: Data transmission takes too long and leads to the violation of timing. [H-3]
IFB-13: Service is requested but not executed correctly. [H-3]
IFB-14: Service is processed with unauthenticated or modified data (e.g. requests from unauthenticated parties). [H-2/4]
IFB-15: Service is processed with incorrect algorithms, which may cause insecure behaviors (e.g. bypassing real processes and reply a faked response). [H-3]
IFB-16: Service is processed with information leakage risks. [H-1]
IFB-17: Service process takes too long and leads to violation of timing. [H-3]

[1] VI represents the vehicle interface.
[2] ED represents the end device.

Then, LSs are discussed, which describes causal factors that may lead to IFBs and cause losses. 23 general loss scenarios are identified in Table 6 with identification instructors.

Finally, system constraints are derived in two ways. The one is simply to invert LSs, and the other is to define system reactions when LSs occur to minimize losses [9]. For example, the system constraint from LS-1 is 'SC-1: the data check must not be bypassed' and 'SC-2: if the check is bypassed, it must be detected and aborted'. LS-1 also explores a system vulnerability that the check process has the risk of being bypassed. The system constraints, as the outcomes of the security analysis phase, help designers pay more attention to system weaknesses.

Table 6 LSs of the Target System

| IFB Index | Itself | Environment ||||
|---|---|---|---|---|---|
| | Algorithm | Input | Calling Behavior | Computing Resource | On Link |
| IFB-1 | / | / | LS-1 | / | / |
| IFB-2 | LS-2 | / | / | / | / |
| IFB-3 | LS-3 | / | / | / | / |
| IFB-4 | LS-4 | / | / | LS-5 | / |
| IFB-5 | / | LS-6 | / | / | / |
| IFB-6 | LS-7 | / | / | / | / |
| IFB-7 | LS-8 | / | / | LS-9 | / |
| IFB-8 | / | / | / | / | LS-10 |
| IFB-9 | / | LS-11 | / | / | / |
| IFB-10 | / | / | / | / | LS-12 |
| IFB-11 | LS-13 | / | / | / | LS-14 |
| IFB-12 | / | / | / | / | LS-15 |
| IFB-13 | LS-16 | LS-17 | / | / | LS-18 |
| IFB-14 | / | LS-19 | / | / | / |
| IFB-15 | LS-20 | / | / | / | / |
| IFB-16 | LS-21 | / | / | / | / |
| IFB-17 | LS-22 | / | / | LS-23 | / |
| LS-n Description ||||||

LS-1: Data check is bypassed, and a fake OK result is output.
LS-2: No or inadequate algorithm is used to check the authenticity and integrity of the data.
LS-3: The algorithm is modified, and a fake NOK is output.
LS-4: The algorithm is modified and requires more computing resources.
LS-5: The adversary occupies computing resources and makes it not enough for the target function.
LS-6: Unauthenticated or modified data are not detected before and input to transform.
LS-7: The algorithm is modified for malicious purposes.
LS-8: The algorithm is modified and requires more computing resources.
LS-9: The adversary occupies computing resources and makes it not enough for the function.
LS-10: Transmission is interrupted intendedly by causing errors on the networks (e.g. broken or shorten links).
LS-11: Unauthenticated or modified data are not detected before and input to transmit.
LS-12: Data is modified when transmitting on links (e.g. man-in-the-middle attack).
LS-13: No or inadequate anti-leakage algorithm is used for data transmission.
LS-14: Links are not protected, the adversary can get access to data on links.
LS-15: Transmission is slowed down by additional mechanisms on links (e.g. additional switches).
LS-16: The algorithm is modified to reject legal requests.
LS-17: Modified system states are input, which makes the system think that the service pre-conditions are not met.
LS-18: Service requests are blocked on links.
LS-19: Unauthenticated or modified data are not detected and input as requested service data.
LS-20: The algorithm is modified for malicious purposes.
LS-21: No or inadequate anti-leakage algorithm is used for data transmission.
LS-22: The algorithm is modified and requires more computing resources.
LS-23: The adversary occupies computing resources and makes it not enough for the function.

## 4.3. Concept Traceability

All concepts in the proposed approach are traceable and linked with each other, which helps to review the security analysis and provide evidence for a proper design process. Referring to the traceability diagram in [9], the diagram of the proposed approach is shown in Figure 5.

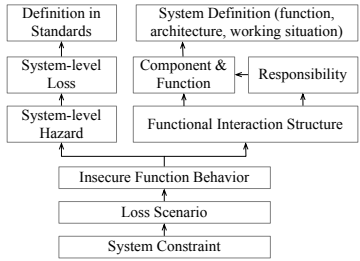

Figure 5 Traceability of the Proposed Approach

## 4.4. An Example Case

Section 4.2 presents the guideline for analyzing in-vehicle network systems. For designing a particular system, a more concrete analysis should be conducted. This section shows how to use the guideline to achieve refined results.

Firstly, a particular use case needs to be defined. Figure 6 shows a simplified domain-based in-vehicle communication system, whose purpose is to update software of the Domain Control Unit (DCU) #1. This use case is assumed to be synchronous type and happen at manufacturer places, which means the attacker may be the insider and have basic knowledge and time to perform attacks. The connection type has no impact on this analysis and are is not care.

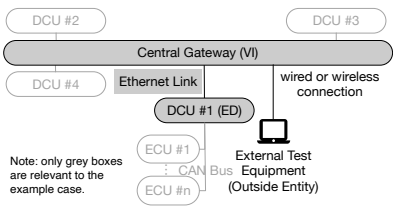

Figure 6 System Architecture of the Example Use Case

Since the first step is performed at a high system level, losses and hazards in particular cases are normally the same as listed in the guideline (in Table 2 and Table 3). This particular analysis starts from Step 2, in which concrete functions of each component are listed with corresponding IFBs in Table 7.

Table 7 Functions (F) with IFBs of the Example Case

| F in VI | F on Link | F in ED | NPCH | PCH | TI |
|---|---|---|---|---|---|
| F-1 |  | F-4 | IFB-1 | IFB-2/3 | IFB-4 |
| F-2 |  | F-5 | / | IFB-5/6 | IFB-7 |
|  | F-3 |  | IFB-8 | IFB-9/ 10 | IFB-11 |
|  |  | F-6 | IFB-12 | IFB-13/14/15/ 16 | IFB-17 |
| Functions Description |||||||

F-1: Check Data function in VI
F-2: Transform Data function in VI
F-3: Transmit Data function on links
F-4: Check Data function in ED
F-5: De/Encapsulate Requests/Responses function in ED
F-6: Process Service function in ED

Then, function-related LSs are identified and described with more details. Some example LSs for F-1 are shown in Table 8.

Table 8 Example Loss Scenarios

| F | IFB | LS |
|---|---|---|
| F-1: Check Data Function in VI | F-1_IFB-2.1: Modified data are not detected and output an OK result. [H-2] | F-1_IFB-2.1_LS-2.1: No algorithm is used to check data integrity. The modified data are faked to be correct data and transmitted into the system. |
| | | F-1_IFB-2.1_LS-2.2: The existing algorithm for data integrity check is not good enough. The well-faked input data can cheat the system and output an 'ok' result. |
| | | F-1_IFB-2.1_LS-2.3: The original vehicle interface is replaced by a fake interface with no or broken algorithm of the data integrity check, which let the modified data be transmitted into the system. |

Finally, the system constraints (shown in Table 9) are derived from LSs with traceable labels.

Table 9 Example System Constraints

| Index | System Constraint | Linked LS |
|---|---|---|
| SC-1 | Adequate algorithm for data integrity check must be used in the vehicle interface. | F-1_IFB-2.1_LS-2.1/2.2 |
| SC-2 | The vehicle interface must be protected from being replaced by a fake interface. | F-1_IFB-2.1_LS-2.3 |
| SC-3 | If the modified data passes the check in the vehicle interface, it must be detected and recorded anywhere else, and the service must be aborted. | F-1_IFB-2.1_LS-2.1/2.2 |
| SC-4 | If the vehicle interface has been replaced by a fake one, it must be detected. The system must reject providing services with insecure components. | F-1_IFB-2.1_LS-2.3 |

## 5. Discussion

To find differences between the system-oriented approach and the threat-oriented one, we compared the presented guideline and the analysis outcomes by using the EVITA threat-oriented method. The attack tree of an OBD flashing use case was reported with five identified attack objects and ten attack methods [4]. By comparing, we found that all attack methods in the EVITA can be mapped to one or more similar LSs in our guideline. By contrast, some kinds of LSs like LS-1 and LS-7 cannot be interpreted by EVITA outputs. Besides, the proposed approach also reveals relationships between concepts with traceable analysis paths (see Figure 5), while the EVITA one uses brainstorms based on expert knowledge and hard to find logical links between identified objects or attack methods.

The most significant feature of this STPA-based approach is the paradigm shift from focusing on threats to system vulnerabilities. The external threats cannot be controlled by the designer but the internal system weaknesses can, which makes it more efficient to protect the system from known or even unknown threats. Besides, several STPA extensions have been proposed in different fields. Therefore, the proposed approach can be well integrated with other STPA-based approaches and perform co-analysis in multi-disciplines. Furthermore, STPA has been proven to be useful at analyzing high system levels, therefore using a similar framework during the whole system design, from the high level to the lower one, would make the work process simpler and outcomes could have strong links with each other.

The general idea of this research is to use the optimized STPA-based approach to analyze security issues of a sort of systems, which is the in-vehicle diagnostic and software update system, and provide a security analysis guideline for this class. The presented guideline with instructors can guide the analysis of a real system in practice. The outline of the research idea is shown in Figure 7.

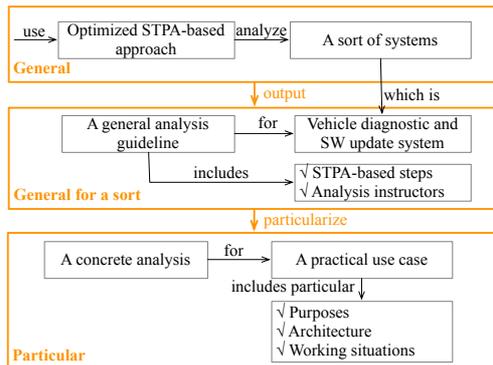

Figure 7 Research Outline

## 5. Conclusion and Future Work

In this paper, we proposed a system-oriented approach for security analysis of in-vehicle communication systems and a guideline for diagnostic and software update use cases. Current research states were introduced, and an example was presented to show how to use the approach. Finally, we conducted a comparison between system- and threat-oriented analyses, highlighted the merits of the proposed approach and presented the general research outline.

Two works have been planned for future researches. Firstly, more particular analyses of industry cases will be performed to exploit approach weaknesses in practice and make it more operable and suitable for real situations. Secondly, we will use this approach to analyze other information-based systems in vehicles, like infotainment system and Advanced Driver Assistance Systems (ADAS) to extend the applicability of the approach and also propose security design guideline for other key systems in vehicles.

## Acknowledgement

This research was supported by the China Scholarship Council and funds of the German Federal Ministry of Education and Research under grant number 16KIS0995. The responsibility for the content of this publication lies with the authors.